\documentclass[aps,twocolumn,prd,showpacs,nofootinbib]{revtex4}

\usepackage{amsmath}
\usepackage{graphicx}
\usepackage{dcolumn}
\usepackage{bm}
\usepackage{amssymb}
\usepackage{latexsym}

\def\be{\begin{equation}}
\def\ee{\end{equation}}
\def\ba{\begin{eqnarray}}
\def\ea{\end{eqnarray}}

\begin{document}

\title{ Phase Diagram for Nflation}

\author{Iftikhar Ahmad$^{a}$\thanks{e-mail:iftikharwah@gmail.com},
Yun-Song Piao$^{a}$\thanks{e-mail:yspiao@gucas.an.cn} and
Cong-Feng Qiao$^{a,b}$\thanks{e-mail:qiaocf@gucas.an.cn}\\
\textit{$a)$College of Physical Sciences,Graduate University of
CAS, YuQuan Road 19A, Beijing 100049, China\\
$b)$ Theoretical Physics Center for Science Facilities (TPCSF),
CAS., Beijing 100049, China}}

\begin{abstract}
Recently, it was showed that there is a large N phase transition
in Nflation, in which when the number of fields is large enough,
the slow roll inflation phase will disappear. In this brief
report, we illustrate the phase diagram for Nflation, and discuss
the entropy bound and some relevant results. It is found that near
the critical point the number of fields saturates dS entropy.


\end{abstract}
\maketitle

\section{Introduction}

Recently, Dimopoulos et.al \cite{DKMW1} showed that the many
massive axion fields predicted by string vacuum can be combined
and lead to a radiatively stable inflation, called Nflation, which
is an interesting implement of assisted inflation mechanism
proposed by Liddle et.al \cite{LMS1}, see also Refs. \cite{MW1,
Piao02061} for many studies, and a feasible embedding of inflation
in string theory. Then Easther and McAllister found \cite{EM1}
that for the mass distribution following Mar$\check{c}$enko-Pastur
law, the spectral index of scalar perturbation is always redder
than that of its corresponding single field. However, this result
is actually valid for any mass contribution and initial condition
of fields, as has been shown in \cite{KL1, SA1} numerically and in
\cite{Piao} analytically. In addition, it was found for Nflation
that the ratio of tensor to scalar is always same as in the single
field case \cite{AL1} and the non-Gaussianity is small \cite{SA21,
BB1}, see also Refs.\cite{SL1} for relevant studies.



In inflation, when the value of field increases up to some value,
the quantum fluctuation of field will be expected to overwhelm its
classical evolution. In this case, the inflaton field will undergo
a kind of random walk, which will lead to the production of many
new regions with different energy densities. This was called as
eternal inflation \cite{V1983a, L1986a}. In principle, it was
thought that dependent on the value of field, there are generally
three different phases in single field inflation, i.e. eternal
inflation phase, slow roll inflation phase and fast roll phase,
which should be also valid for Nflation.

However, recently, it was found \cite{APQ2} that when the number
of fields is large enough, the slow roll inflation phase will
disappear, which means there exists a large N transition for
Nflation. The reason is, though the end value of slow roll
inflation decreases with the increase of number $N$ of fields, the
value separating the slow roll inflation phase and the eternal
inflation phase, hereafter called as the eternal inflation
boundary for convenience, decreases more rapidly, thus they will
cross inevitably at some value of $N$, after this the slow roll
inflation phase will go out of sight. This result means there is a
bound for the number of fields driving the slow roll Nflation.
This is also consistent with recent arguments from black hole
physics \cite{D07, D08}, in which there exists a gravitational cutoff,
whose value equals to our bound, beyond which the quantum gravity
effect will become important, see also Refs. \cite{Huang, LS1} for
some similar bounds.

In single field inflation, when the inflaton field is in its
eternal inflation boundary the primordial density perturbation
${\delta\rho}/\rho \sim 1$, thus it will be hardly possible for us
to receive the information from the eternal inflation phase, since
in that time we will be swallowed by black hole \cite{RBI}. This
result may be actually assured by a relation between the entropy
and the total efolding number \cite{NSAEG}, in which when
${\delta\rho}/\rho \sim 1$, the entropy in unit of efolding number
is less than one, which means we can not obtain any information.
Thus it is significant to examine how above results change for
Nflation, especially what occurs around its phase transition
point. It can be expected that there maybe more general and
interesting results. In this paper, we will firstly illustrate the
phase diagram for Nflation, and then give relevant discussions.

\section {Phase diagram for Nflation}

In the Nflation model, the inflation is driven by many massive
fields. For simplicity, we assume that the masses of all fields
are equal, i.e. $m_i=m$, and also $\phi_i=\phi$, which will also
be implemented in next section. Following Ref. \cite{APQ2}, the
end value of slow roll inflation phase and the eternal inflation
boundary with respect to $N$ are given by \be \phi\simeq
{M_p\over\sqrt {N}}, \label{ep}\ee \be \phi \simeq {1\over {
N}^{3/4}}\sqrt{M_p^3\over m}, \label{cr}\ee respectively. It can
be noticed that the end value goes along with ${1\over\sqrt {N}}$,
it decreases slower than the eternal inflation boundary with $N$,
since the latter changes with ${1\over { N}^{3/4}}$. Thus when we
plot the lines of the end value and the eternal inflation boundary
moving with respect to $N$, respectively, there must be a point
where these two lines cross, see Fig.1. This crossing point is \be
{N}\simeq {M_p^2\over m^2}, \label{caln}\ee beyond which the slow
roll inflation phase will disappear. Thus here we call this point
as the critical point. It seems be expected that after the
critical point is got across, the line denoting the eternal
inflation boundary will not extend downwards any more, the line
left is that denoting the end value, which still obeys
Eq.(\ref{ep}), see the dashed line of Fig.1. The reason is the
calculation of the eternal inflation boundary is based on the slow
roll approximation, while below the end value the slow rolling of
field is actually replaced by the fast rolling, in this case the
quantum fluctuation is actually suppressed, thus it is hardly
possible that the quantum fluctuation of field will overwhelm its
classical evolution. However, the case maybe not so simple. In
next section, we will see there is an entropy bound for the number
of fields, and at the critical point this bound is saturated. This
means that beyond the critical point our above semiclassical
arguments can not be applied. Thus in this sense in principle what
is the diagram beyond the critical point remains open.

\begin{figure}[t]
\begin{center}
\includegraphics[width=8cm]{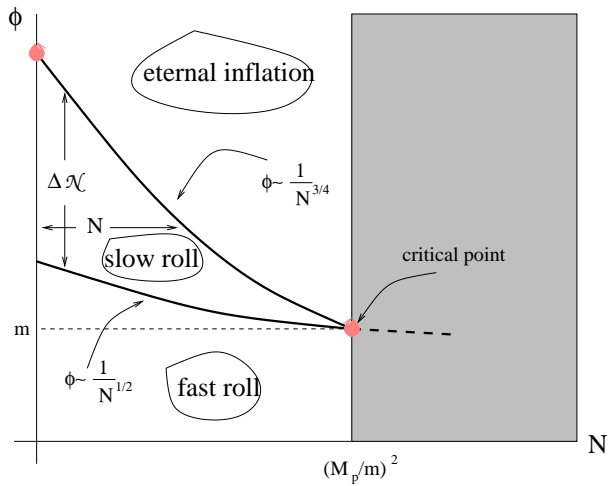}
\caption{ The $\phi-N$ phase diagram for Nflation. The upper solid
line is the eternal inflation boundary and the lower solid line is
the end value of the slow roll inflation. These two lines split
the region into three phases, i.e. eternal inflation phase, slow
roll inflation phase and fast roll phase. There is a critical
point, beyond which the slow roll inflation phase disappears.
 }\label{xx}
\end{center}
\end{figure}

The value of fields at critical point can be obtained by
substituting Eq.(\ref{caln}) into any one of Eqs.(\ref{ep}) and
(\ref{cr}), which is $\phi\simeq m$. This indicates that if
initially $\phi< m$, no matter what ${N}$ is, the slow roll
inflation will not occur. The existence of slow roll inflation is
important for solving the problems of standard cosmology and
generating the primordial perturbation seeding large scale
structures. In the phase diagram Fig.1, we can see that the slow
roll inflation phase is in a limited region, which means in order
to make Nflation responsible for our observable universe, the
relevant parameters must be placed suitably.

We assume that all mass are equal only for simplicity. For the
case that not all mass are equal, the result is also similar, as
has been shown in Ref. \cite{APQ2}, in which the mass distribution
following Mar$\check{c}$enko-Pastur law \cite{EM1} is taken for
calculations. Thus the phase diagram is still Fig.1, the only
slight difference is replacing $m$ with the average mass $\bar m$.
It should be noted that here in the phase diagram the number $N$
of fields dose not include massless scalar fields. The reason is
when the masses of fields are negligible, they will not affect the
motion of massive fields dominating the evolution of universe,
while the perturbations used to calculate the quantum jump of
fields are those along the trajectory of fields space, since the
massless fields only provide the entropy perturbations orthogonal
to the trajectory, which thus are not considered in the
calculations deducing Eqs.(\ref{ep}) and (\ref{cr}). Thus if there
are some nearly massless fields and some massive fields with
nearly same order, it should be that there is a bound $N\lesssim
M_p^2/{\bar m}^2$, in which only massive fields are included in
the definition of $\bar m$ and $N$.

\section{Discussion}

\subsection{On primordial density perturbation at the eternal
inflation boundary}

In single field inflation, when the inflaton field is in its
eternal inflation boundary, the primordial perturbation
${\delta\rho}/\rho \sim 1$. The primordial density perturbation
during Nflation can be calculated by using the formula of Sasaki
and Stewart \cite{SS}. In slow roll approximation,
$\left(\frac{\delta\rho}{\rho}\right)^2 \sim {m^2{ N}^2\phi^4\over
M_p^6}$ \cite{KL1, LR1}. The motion of the eternal inflation
boundary obeys Eq.(\ref{cr}). Thus substituting Eq.(\ref{cr}) into
it and then cancelling the variable $\phi$, we can obtain
\ba{\delta\rho\over \rho}\simeq \frac{1}{\sqrt{ N}},\label{a13}\ea
where the factor with order one has been neglected, which
hereafter will also be implemented. We can see that
${\delta\rho}/{\rho}$ is decreased with respect to the increase of
$N$, and for each value of $N$, ${\delta\rho}/{\rho}$ is always
less than one. This result is obviously different from that of
single field. The reason leading to this result is, in single
field inflation the eternal inflation boundary and the point that
the density perturbation equals to one are same, however, the
changes of both with $N$ are different, one is $\sim 1/{N}^{3/4}$
and the other is $\sim 1/\sqrt{N}$. Intuitively, the eternal
inflation means that the quantum fluctuations of fields lead to
the production of many new regions with different energy
densities, thus it seems that when we approach the eternal
inflation boundary the density perturbation will be expected to
near one. Thus in this sense our result looks like unintuitive.
However, in fact what the eternal inflation phase means should be
a phase in which the quantum fluctuation of field overwhelms its
classical evolution, which is not certain to suggest that the
density perturbation is about one.

Thus different from single field inflation, in which we are
impossible to receive the information from the eternal inflation
phase since in that time the black hole has swallowed us due to
the primordial density perturbation with near one, it seems that
when $N$ is large, we may obtain some information from the eternal
inflation phase, at least in principle we can obtain those from
the boundary of eternal inflation phase. Beyond this boundary, the
fields are walked randomly, thus the slow roll approximation is
broken and the results based on the slow roll approximation are
not robust any more. In principle, for the eternal inflation phase
of Nflation we need to calculate the density perturbation in a new
way to know how much it is actually, which, however, has been
beyond our capability. The eternal inflation phase for single
fields has been studied by using the stochastic approach
\cite{AAS}.


\subsection{ On entropy bound}

The entropy during Nflation can be approximately given by dS
entropy $S\sim {M_p^2\over H^2}$. Here we regard $S$ as the
entropy at the eternal inflation boundary. Thus we have \be S \sim
{M_p^2\over H^2}\sim {M_p^4\over Nm^2\phi^2}\sim \sqrt{N}{M_p\over
m}, \label{s}\ee where Eq.(\ref{cr}) has been used. It is interesting to find
that $S$ is proportional to $\sqrt{N}$, which means the entropy increases
with the number of field. Here the case is slightly similar to that of
the entanglement entropy for a black hole, in which there seems be a
dependence of the entanglement entropy on $N$, which conflicts the usual
result of black hole entropy,
since each of fields equally contributes to the entropy \cite{D08}.
However, this problem may be solved by invoking the correct gravity cutoff
$\Lambda\sim {M_p\over\sqrt{N}}$ \cite{D07}, as has been argued in
Ref. \cite{D08}. In Eq.(\ref{s}) if we replace $M_p$
with a same gravity cutoff $\Lambda$, then we will obtain $S\sim \sqrt{N}
{\Lambda\over m}\sim {M_p\over m}$, which is just the result for single
field, i.e. $S\sim {M_p\over m}$ at the eternal inflation boundary. Thus
it seems that the argument in Ref. \cite{D08} is universal for the
relevant issues involving $N$ species.

It can be noticed that
the efolding number ${\cal N}\sim {N\phi^2\over M_p^2}$. For
initial $\phi$ being in its eternal inflation boundary, where
$\phi$ is given by Eq.(\ref{cr}), for fixed $N$, i.e. along the
line paralleling the $\phi$ axis in Fig.1, $\cal N$ obtained will
be the total efolding number along corresponding line in slow roll
inflation phase, hereafter called $\Delta {\cal N}$, see Fig.1.
Thus with Eq.(\ref{cr}), we can have $\Delta {\cal N}\sim
{M_p\over m \sqrt{N}}$. Then we substitute it into Eq.(\ref{s}),
and thus for the eternal inflation boundary, we have \be N\cdot
\Delta {\cal N}\simeq S, \label{s11}\ee which is a general entropy
bound including $N$, and is also our main result. It means that
below the eternal inflation boundary, we have the bound $N\cdot
\Delta {\cal N} \lesssim S $. This result indicates that for fixed
$N$, i.e. along the line paralleling the $\phi$ axis in Fig.1, the
total efolding number $\Delta {\cal N}$ of slow roll inflation
phase is bounded by $S$, while for fixed $\Delta {\cal N}$, i.e.
along the line paralleling the $N$ axis in Fig.1, the number $N$
of fields is bounded by $S$, and at the eternal inflation
boundary, the entropy bound is saturated.

There are two special cases, corresponding to the regions around
red points in Fig.1. For details, one is that for $N=1$, i.e.
single field, we have $\Delta {\cal N}\simeq S$ from
Eq.(\ref{s11}), thus the result for single field is recovered
\cite{NSAEG}. Following \cite{NSAEG} to large N, Eq.(\ref{s11})
can be actually also deduced. By making the derivatives of $\cal
N$ and $S$ with respect to the time, respectively, we can have
\ba\frac{d{\cal N}}{dS}\simeq \frac{ M_p^2}{m^2S^2}, \label{a11}
\ea where $S$ is the function of $\phi$, see the second equation
in Eq.(\ref{s}), and thus can be used to cancel $\phi$. By
integrating this equation along the line paralleling the $\phi$
axis in Fig.1, where the lower limit is the eternal inflation
boundary and the upper limit is the end value of slow roll
inflation phase, and then applying approximation condition
$\phi_e\ll\phi$, where $\phi$ and $\phi_e$ represent the values of
eternal inflation boundary and the end of slow roll inflation,
respectively, which actually implies that $S_e\gg S$ and thus
$({S_e-S})/{S_e}\simeq 1$, we have \ba \Delta {\cal N}\simeq
(\frac{\delta\rho}{\rho})^2S,  \label{a12}\ea where
$\left(\frac{\delta\rho}{\rho}\right)^2\sim { M_p^2\over m^2 S^2}$
has been applied, which can be obtained since both
${\delta\rho\over \rho}$ and $S$ are the functions of $\phi$. This
result has been showed in Ref. \cite{NSAEG} for single field,
however, since Eq.(\ref{a12}) is independent on the number $N$ of
fields, thus it is still valid for $N$ fields. For single field
inflation, ${\delta\rho\over \rho}\sim 1$ only at eternal
inflation boundary, thus we always have $\Delta {\cal N}\lesssim
S$ for slow roll phase, i.e. the total efolding number is bounded
by the entropy, which is saturated at eternal inflation boundary.
Note that Eq.(\ref{a12}) is an integral result in which
$\delta\rho/\rho $ with the change of $\phi$ and thus $S$ is
condidered, which is slightly different from that in Ref.
\cite{HLW}. Thus combining Eqs.(\ref{a13}) and (\ref{a12}), we can
find Eq.(\ref{s11}) again. This also indicates the result of
Eq.(\ref{a13}) is reliable.

The other is that for $N$ being near its critical point, in which
approximately we have $\Delta {\cal N}\simeq 1$, thus we can
obtain $N\simeq S$, i.e. $S$ is saturated by the number $N$ of
fields. This can also be seen by combining Eq.(\ref{caln}) for the
critical point and Eq.(\ref{s}), in which we can find $S\simeq N $
at the critical point.

Thus below the critical point, $ N\lesssim S$.
From Eq.(\ref{s}), $S\sim \sqrt{N}{M_p\over m} \gtrsim
N$ can be obtained. This means $N \lesssim ({M_p\over m})^2$.
In Refs. \cite{D07},
it was argued that $M_p$ is renormalized in the presence of $N$
fields at scale $m$ so that $M_p^2\gtrsim Nm^2$, in other words,
$N>({M_p\over m})^2$ is inconsistent. Here, if $N>({M_p\over m})^2$,
then combining it and Eq.(\ref{s}), we will have $N>S$,
i.e. the number $N$ of fields is larger than the
dS entropy of critical point. This is certainly impossible,
since intuitively it may be thought that there is at least a
freedom degree for each field, thus the total freedom degree of
$N$ fields system, i.e. the entropy, should be at least $N$, while
dS entropy is the maximal entropy of a system. Thus we arrive at
same conclusion with Ref. \cite{D07} from a different viewpoint.
This again shows the consistence of our result.


\textbf{Acknowledgments} We thank Y.F. Cai for helpful discussions
and comment. I.A thanks the support of (HEC) Pakistan. This work
is supported in part by NSFC under Grant No: 10491306, 10521003,
10775179, 10405029, 10775180, in part by the Scientific Research
Fund of GUCAS(NO.055101BM03), in part by CAS under Grant No:
KJCX3-SYW-N2.

\end{document}